\documentclass[12pt,aoas, noinfoline]{imsart}
\usepackage{lipsum}
\usepackage{graphicx}
\usepackage[colorlinks, allcolors=NavyBlue]{hyperref}
\usepackage[dvipsnames]{xcolor}
\usepackage[nameinlink, noabbrev]{cleveref}
\usepackage{natbib}
\usepackage{caption, subcaption}
\usepackage{slashbox}
\usepackage{tikz}
\usepackage{amsmath, amsfonts}
\usepackage{cuted}
\usepackage{lipsum}
\usepackage{floatrow}
\usepackage{bm}
\usepackage{amsthm}
\usepackage{algorithm}
\usepackage{algorithmicx, algpseudocode}
\usepackage{multirow}
\usepackage{adjustbox}

\usepackage{tcolorbox}
\definecolor{mycolor}{rgb}{0.122, 0.435, 0.698}
\newtcbox{\mybox}{nobeforeafter,colframe=mycolor,colback=mycolor!10!white,boxrule=0.5pt,arc=4pt,
  boxsep=0pt,left=2pt,right=2pt,top=2pt,bottom=2pt,tcbox raise base}
\newtcbox{\myboxx}{nobeforeafter,colframe=orange,colback=orange!10!white,boxrule=0.5pt,arc=4pt,
  boxsep=0pt,left=2pt,right=2pt,top=2pt,bottom=2pt,tcbox raise base}

\usetikzlibrary{arrows,shapes,trees, positioning}
\def \Theta{45}
\def \dist{2.5cm}

\def \stubup{0.9}
\def \stubside{ {\stubup / sin(\Theta)} }

\def \radsmall{1.2}
\def \radbig{2.4}

\theoremstyle{remark}
\newtheorem{example}{Example}

\renewcommand{\P}{\mathbb{P}}
\newcommand{\E}{\mathbb{E}}
\newcommand{\R}{\mathbb{R}}
\renewcommand{\epsilon}{\varepsilon}

\newcommand\Tstrut{\rule{0pt}{2.6ex}}         

\begin{document}

\begin{frontmatter}
\title{A Goodness-of-Fit Test for Sampled Subgraphs}
\runtitle{A Goodness-of-Fit Test for Sampled Subgraphs}
\thankstext{T1}{I'd like to thank Firmin Doko Tchatoka, Simon Fielke, Petra Kuhnert, and Virginie Masson for their helpful comments.} 

\begin{aug}
  \author{\fnms{Robert}  \snm{Garrard}\corref{}\ead[label=e1]{rob.garrard@csiro.au}},
  \runauthor{R. Garrard}
  \affiliation{Commonwealth Scientific and Industrial Research Organisation (CSIRO)}
  \address{41 Boggo Rd, Dutton Park, QLD, 4102, Australia,\\
  \printead{e1}}
\end{aug}

\begin{abstract}
We consider the problem of testing whether a graph's degree distribution belongs to a particular family, such as poisson or scale-free, given that we only observe a sampled subgraph. In particular, we focus on induced subgraph sampling, a sampling design which systematically distorts the degree distribution of interest. We estimate the parameter indexing the hypothesized family by generalized method of moments and utilize the Kolmogorov-Smirnov test statistic to assess goodness-of-fit. Since the distribution in the null hypothesis has been estimated, critical values for the test statistic must be simulated. We propose a novel bootstrap in which we construct a graph whose degree distribution conforms to the null hypothesis from which we may draw pseudo-samples in the form of induced subgraphs. We investigate the properties of this procedure with a monte carlo study which confirms that the bootstrap is able to attain size close to the nominal level while exhibiting power under the alternative hypothesis. We present an application of this test to the protein interaction network (PIN) of the yeast \emph{Saccharomyces cerevisiae}. Accounting for the high rates of false negatives present in PIN measurement, we are able to reject the hypothesis that the PIN of S. \emph{cerevisiae} follows an Erd\H{o}s-R\'{e}nyi random graph family of degree distributions.
\end{abstract}



\end{frontmatter}

\section{Introduction}

The degree distribution is one of the fundamental descriptions of a graph, often leading to models of network formation being referred to by their resulting degree distributions \citep{Newman2001}. Scale-free graphs, whose degree distributions obey a power law, are especially interesting as they exhibit a ``robust-yet-fragile'' property in which the graph is resilient to the failure of randomly selected vertices but vulnerable to catastrophic failure in the event that its high-degree hubs are targeted \citep{Albert2000, Doyle2005}. This feature is common to many real-world graphs, such as interbank lending networks and the webgraph of the internet \citep{Boss2004, Gai2011, Doyle2005}. 

We consider the problem of testing the hypothesis that a graph's degree distribution belongs to a particular family, such as poisson or power law, based on a sampled subgraph but without requiring that the hypothesized distribution be fully parameterized. This is akin to \citet{Lilliefors1967, Lilliefors1969}. Supposing that an i.i.d sample of degrees were acquired, this would be a standard goodness-of-fit test with critical values computable with the bootstrap.\footnote{In the case of scale-free distributions, critical values for the Kolmogorov-Smirnov statistic have been tabulated by \citet{Goldstein2004}}However, graphs are often sampled in a fashion where observed degrees may not be considered i.i.d draws from the population \citep[Chapter 5]{Kolaczyk2009}. 

\begin{figure}[t]
\begin{subfigure}[b]{0.45\textwidth}
	\quad
	\begin{tikzpicture}
	\node[circle,draw,black, fill=white, line width = 0.33mm] (A) at (90:\radsmall ) {};
	\node[circle,draw,black, fill=blue, line width = 0.33mm] (B) at (90+1*360/5:\radsmall ) {};
	\node[circle,draw,black, fill=white, line width = 0.33mm] (C) at (90+2*360/5:\radsmall ) {};
	\node[circle,draw,black, fill=blue, line width = 0.33mm] (D) at (90+3*360/5:\radsmall ) {};
	\node[circle,draw,black, fill=blue, line width = 0.33mm] (E) at (90+4*360/5:\radsmall ) {};

	\node[circle,draw,black, fill=blue, line width = 0.33mm] (a) at (90:\radbig ) {};
	\node[circle,draw,black, fill=blue, line width = 0.33mm] (b) at (90+1*360/5:\radbig ) {};
	\node[circle,draw,black, fill=white, line width = 0.33mm] (c) at (90+2*360/5:\radbig ) {};
	\node[circle,draw,black, fill=white, line width = 0.33mm] (d) at (90+3*360/5:\radbig ) {};	
	\node[circle,draw,black, fill=blue, line width = 0.33mm] (e) at (90+4*360/5:\radbig ) {};

	\draw[color=black, line width = 0.75] (a) -- (b);
	\draw[color=black, line width = 0.75] (b) -- (c);
	\draw[color=black, line width = 0.75] (c) -- (d);
	\draw[color=black, line width = 0.75] (d) -- (e);
	\draw[color=black, line width = 0.75] (e) -- (a);

	\draw[color=black, line width = 0.75] (a) -- (A);
	\draw[color=black, line width = 0.75] (b) -- (B);
	\draw[color=black, line width = 0.75] (c) -- (C);
	\draw[color=black, line width = 0.75] (d) -- (D);
	\draw[color=black, line width = 0.75] (e) -- (E);

	\draw[color=black, line width = 0.75] (A) -- (C);
	\draw[color=black, line width = 0.75] (A) -- (D);
	\draw[color=black, line width = 0.75] (B) -- (D);
	\draw[color=black, line width = 0.75] (B) -- (E);
	\draw[color=black, line width = 0.75] (C) -- (E);
	\end{tikzpicture}
\caption{Population Graph}
\label{fig:inducedsubgrapha}
\end{subfigure}
\hfill
\begin{subfigure}[b]{0.45\textwidth}
	\quad
	\begin{tikzpicture}
	\node[circle,draw,black, fill=blue, line width = 0.33mm] (B) at (90+1*360/5:\radsmall ) {};
	\node[circle,draw,black, fill=blue, line width = 0.33mm] (D) at (90+3*360/5:\radsmall ) {};
	\node[circle,draw,black, fill=blue, line width = 0.33mm] (E) at (90+4*360/5:\radsmall ) {};

	\node[circle,draw,black, fill=blue, line width = 0.33mm] (a) at (90:\radbig ) {};
	\node[circle,draw,black, fill=blue, line width = 0.33mm] (b) at (90+1*360/5:\radbig ) {};
	\node[circle,draw,white, fill=white, line width = 0.33mm] (c) at (90+2*360/5:\radbig ) {};
	\node[circle,draw,white, fill=white, line width = 0.33mm] (d) at (90+3*360/5:\radbig ) {};	
	\node[circle,draw,black, fill=blue, line width = 0.33mm] (e) at (90+4*360/5:\radbig ) {};

	\draw[color=black, line width = 0.75] (a) -- (b);
	\draw[color=black, line width = 0.75] (e) -- (a);

	\draw[color=black, line width = 0.75] (b) -- (B);
	\draw[color=black, line width = 0.75] (e) -- (E);

	\draw[color=black, line width = 0.75] (B) -- (D);
	\draw[color=black, line width = 0.75] (B) -- (E);

	\end{tikzpicture}
\caption{Induced Subgraph}
\label{fig:inducedsubgraphb}
\end{subfigure}
\hfill

\caption{\ref{fig:inducedsubgrapha} shows a graph representing a population of ten vertices where six vertices have been selected through simple random sampling. \ref{fig:inducedsubgraphb} displays the subgraph induced by those vertices.} 
\label{fig:inducedsubgraph}
\end{figure}
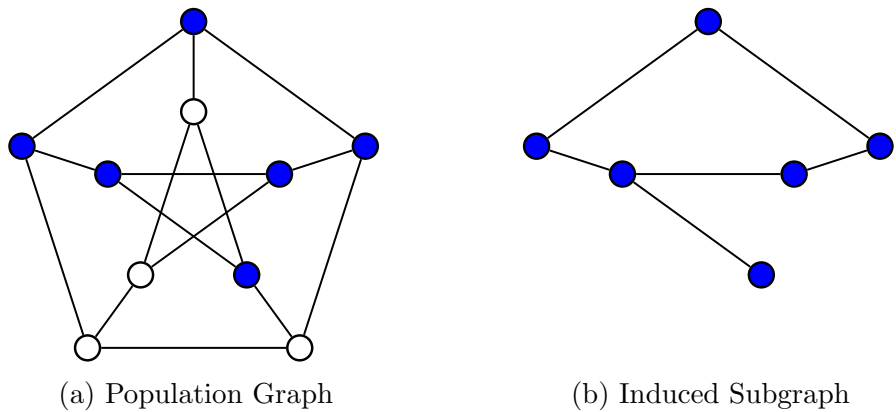

Here we consider samples acquired to be induced subgraphs. Induced subgraph sampling involves randomly sampling a subset of vertices without replacement and retaining only the edges in the population between the vertices sampled. This sampling design systematically distorts the degree distribution as vertices with high degree are more likely to have neighbors included in the sample, and therefore an edge included, than vertices with low degree. \Cref{fig:inducedsubgraph} illustrates this procedure. Nonparametric estimation of the degree distribution based on induced subgraph samples, which has been considered by \citet{Frank1980} and \citet{Zhang2015}, has proven to be particularly challenging.

In this article we propose a novel bootstrap which allows for critical values to be simulated for a goodness-of-fit test when the sample acquired is an induced subgraph. We propose estimating the parameter indexing the hypothesized family of distributions using generalized method of moments (GMM), for which we exploit a moment condition from \citet{Frank1980} and an approximation to its covariance matrix from \citet{Zhang2015}. Once the null distribution has been estimated, we construct a graph whose degree distribution conforms to the null hypothesis from which we may draw pseudo-samples in the form of induced subgraphs. We conduct a monte carlo experiment for poisson and power law data generating processes to assess the size and power properties of the test for various sampling rates. We confirm that the bootstrap is able to control the size of the test close to the nominal level while exhibiting power under most alternatives. However, in some regions of the parameter space the test is unable to distinguish between the two DGPs, particularly for small sampling rates.

We apply this test to the protein interaction network (PIN) of the budding yeast \emph{Saccharomyces cerevisiae}. One of the common methods for measuring PINs, the yeast two-hybrid assay (Y2H), detects pairwise interactions between a set of sampled proteins. Thus, PINs that result from Y2H assays constitute induced subgraph samples. The degree distribution of a PIN is an especially interest feature of the network as the degree of a protein in the network has been noted to correlate with its essentiality for the viability of a cell or organism \citep{Jeong2001}. If the degree distribution of a PIN followed a power law, with many sparsely connected vertices and only a few highly connected, then therapies which target the function of a highly connected protein present a strategy for medical interventions \citep{Vogelstein2000}. If the PIN's degree distribution is more like than of an Erd\H{o}s-R\'{e}nyi random graph, then such a strategy wouldn't be expected to be as successful. 

Y2H assays are prone to significant measurement error, plagued by high rates of both false positives and negatives. \citet{Han2005} show via simulation study that when such measurement error is introduced, subgraphs formed by scale-free and poisson graphs can be practically indistinguishable. In our application we consider five data sets from \citet{Yu2008}, for which each is believed to have high specificity but low sensitivity. We explicitly incorporate various frequencies of type I error into our test and are able to reject the hypothesis that the S. \emph{cerevisiae} PIN follows a poisson distribution for each.


\section{Setup}

Let $G = (V, E)$ be a simple graph with $|V| = N$ vertices. $\{i, j\} \in E$ if there is an undirected edge between vertices $i$ and $j$. The \emph{degree} of vertex $i$, $d_{i}(G) = \sum_{j\not = i} I( \{i, j\} \in E)$, is the number of vertices with which $i$ shares an edge. The \emph{degree distribution} of $G$ is a vector, $P_{G} \in \R^{N}$, containing the frequency with which vertices of a particular degree occur, $P_{G}(k) = \#\{i : d_{i} = k\}/N$, $k = 0,\dots,N-1$. The cumulative distribution function (CDF) may be defined as usual $F_{G}(k) = \sum_{i\leq k} P_{G}(i)$. Let $G^{\prime} = (V^{\prime}, E^{\prime})$ be an induced subgraph formed by random sampling $|V^{\prime}| = n$ vertices and including an edge $\{i,j\} \in E^{\prime}$ if and only if $i,j \in V^{\prime}$. 

Let $F_{0}(\theta)$ be a family of distributions indexed by an unknown parameter vector $\theta$. The objective is to test the following hypothesis.

\begin{equation}
\label{eqn:hypothesis}
H_{0}: F_{G} = F_{0}(\theta) \quad vs. \quad H_{A}: F_{G} \not= F_{0}(\theta)
\end{equation}

	That is, we are looking to perform a goodness-of-fit test against a family of distributions. For example, we may wish to detect whether the population graph $G$ follows one of the common network distributions, such as poisson or a power law. Standard goodness-of-fit tests for which critical values may be tabulated for the test statistics, such as Kolmogorov-Smirnov or Pearson's Chi-square, require that the hypothesised distribution be fully specified. However, we do not have a priori knowledge of the parameter vector $\theta$ and must estimate it from the sampled subgraph $G^{\prime}$. This is akin to \citet{Lilliefors1967, Lilliefors1969} in that critical values must be extracted through monte carlo simulation.
	
An added difficulty lies in the sampling method. Supposing that a random sample of the degrees of vertices in $G$ were acquired, simulation of the critical values would be relatively straightforward. However, the sample consists of an induced subgraph, wherein the degree of vertices is systematically distorted by measuring edges only to other vertices in the sample. This presents an issue both in the simulation of critical values and in the estimation of the parameter $\theta$.

\subsection{Estimation}

Supposing we draw an induced subgraph sample of size $n$ from a population graph on $N$ vertices, it is relatively easy to show that the distortion to the degree distribution obeys the following.\footnote{Conditional on vertex $i$ having $d_{i}(G) = k$, the probability that $i$ has $d_{i}(G^{\prime}) = j$, $j=0,\dots,k$ in the subgraph is hypergeometrically distributed. Marginalizing yields equation \ref{eqn:distortion}.}

\begin{equation}
\label{eqn:distortion}
\E [P_{G^{\prime}}] = \bm{X}P_{G}
\end{equation}

where

\begin{equation}
\label{eqn:X}
\bm{X}_{ij} = \binom{j}{i}\binom{N-1 - j}{n-1-i}\binom{N-1}{n-1}^{-1}
\end{equation}

for $i,j$ such that the binomial coefficients are defined and $\bm{X}_{ij} = 0$ otherwise. This is assuming that a subset of $n$ vertices is selected by simple random sampling. In the case of Bernoulli sampling, where each vertex is selected to be in the sample independently with probability $p$, then the columns of the design matrix would be binomial distributions rather than hypergeometric. 

It's important to note that $\bm{X}$ has dimensions $n \times N$, with $N \gg n$, so direct estimation of $P_{G}$ via least squares is infeasible.\footnote{It may be tempting to apply the Lasso \citep{Tibshirani1996}, since the vector $P_{G}$ is likely to be sparse, but since the columns of $\bm{X}$ are highly correlated, the Lasso will be unable to recover the support of $P_{G}$.} \citet{Frank1980} constructs an unbiased estimator for $P_{G}$ by assuming a known maxmum degree in the population graph, allowing for the matrix $\bm{X}$ to be truncated such that it becomes square. However, this estimator exhibits extreme estimates, far outside the admissible $[0,1]$ interval, and erratic sign switching behavior. \citet{Zhang2015} identify the origin of this pathological behavior as the design matrix being severely ill-conditioned with singular values decaying quickly to zero. They propose stabilizing the inversion of $\bm{X}$ by using Tikhonov regularization and imposing that the estimate be a valid probability mass by constraint. They choose the tuning parameter in the penalty term using a monte carlo version of Stein's unbiased risk estimate \citep{Stein1981, Ramani2008, Eldar2009}. Here we are able to avoid issues with estimating $P_{G}$ directly as long as the vector $\theta$ parameterizing the family of distributions being tested is of dimension much lower than $n$.

\citet{Zhang2015} also show that for moderate sampling rates, the sampling error term in the linear system 

\begin{equation}
\label{eqn:linearsystem}
P_{G^{\prime}} = \bm{X}P_{G} + \bm{\epsilon}
\end{equation}

may be well approximated by $\bm{\varepsilon} \sim N\left(0, \frac{1}{n}\text{diag}(\bm{X}P_{G}) \right)$.\footnote{Specifically, they show that $P_{G^{\prime}}$ is close in total variation distance to poisson, which for moderate sampling rates may be approximated as gaussian.}

Let $P_{0}(\theta)$ be the probability mass corresponding to the hypothesized distribution $F_{0}(\theta)$. In order to estimate $\theta$ from the induced subgraph sample $G^{\prime}$, we exploit the moment condition in equation \ref{eqn:distortion} to construct a GMM estimator.

\begin{equation}
\label{eqn:GMM}
\hat{\theta} = \underset{\theta}{\text{argmin}} \ n (P_{G^{\prime}} - \bm{X}P_{0}(\theta))^{\prime} \mathbf{W} (P_{G^{\prime}} - \bm{X}P_{0}(\theta))
\end{equation}

where $\mathbf{W}$ denotes a weighting matrix. To perform efficient GMM the weighting matrix in \ref{eqn:GMM} should be chosen to be the inverse covariance matrix of the moment condition. Using the approximation from \citet{Zhang2015} ideally we would choose

\begin{equation}
\label{eqn:efficientGMM}
\mathbf{W} = \left[ \text{diag}(\bm{X}P_{0}(\theta) )\right]^{+} 
\end{equation}

where $\mathbf{A}^{+}$ denotes a pseudo-inverse of $\mathbf{A}$. It must be pseudo-inverted since many elements along the diagonal of $diag(\mathbf{X} P_{0}(\theta))$ will be zero. However, this presents an issue when estimating the parameter of scale-free and other heavy tailed distributions. \citet{Goldstein2004} show that estimating a power law parameter by linear fit on a log-log scale leads to inaccurate estimates, with maximum likelihood demonstrating superior properties. We encounter a similar difficulty in that, for power law distributions, the elements of $\mathbf{X}P_{0}(\theta)$ may not decay quickly enough to zero. This causes elements in the weighting matrix to become greatly inflated. Instead we estimate the weighting matrix using equation \ref{eqn:distortion}.

\begin{equation}
\label{eqn:estimatedGMM}
\hat{\mathbf{W}} = \left[ \text{diag}(P_{G^{\prime}} )\right]^{+} 
\end{equation}

This also guarantees that the program is convex and can be readily solved by numerical algorithms. 

We can now construct our preferred test statistic for assessing goodness of fit. Common choices are the Kolmogorov-Smirnov (KS) statistic \citep{Massey1951} or a statistic of squared contrasts whose null distribution is chi-square \citep{Pearson1900, Wald1943}. Letting $\tilde{F}_{0}(\hat{\theta})$ be the CDF of $\E[P_{G^{\prime}} | \ \hat{\theta}] = \bm{X} P_{0}(\hat{\theta})$, the KS statistic has the form

\begin{equation}
\label{eqn:KS}
D(G^{\prime}) = \sqrt{n} \sup | F_{G^{\prime}} - \tilde{F}_{0}(\hat{\theta})|
\end{equation}

The Wald statistic may be found by substituting $\hat{\theta}$ into the objective function in equation \ref{eqn:GMM}. In what follows we focus on the KS statistic.

\section{Bootstrap}
Here we propose a bootstrap \citep{Efron1979} for simulating the critical values corresponding to the previously constructed statistic $D(G^{\prime})$. A typical parametric bootstrap for a Lilliefors-type test presented here would involve estimating the parameter of the hypothesized distribution, $\hat{\theta}$, drawing many pseudo-samples from the distribtution indexed by that estimate, $F_{0}(\hat{\theta})$, and constructing a bootstrap statistic $D^{*}$ for each pseudo-sample. The distribution of these bootstrap statistics may then be used to calculate P-values or critical values. The departure from standard practice here is that our sample is not an iid draw from $F_{0}$ under the null, rather it is an induced subgraph formed by randomly sampling the vertices of a graph whose degree distribution conforms to $F_{0}$. If we wish for the bootstrap to appropriately immitate the sampling process under the null, we will require a graph with degree distribution $F_{0}(\hat{\theta})$ for arbitrary $\hat{\theta}$. In order to construct a graph conforming to the null hypothesis we employ an algorithm for constructing a graph with a prescribed degree distribution due to the Configuration Model of network formation \citep{Bender1978, Bollobas1982}.

Given a graph, $G_{0}$ on $N$ vertices whose degree distribution conforms to the null hypothesis, $F_{0}(\hat{\theta})$, we suggest the following bootstrap for the Kolmogorov-Smirnov statistic in equation \ref{eqn:KS}.

\begin{algorithm}[H]
\caption{Graphical Bootstrap}
\label{alg:bootstrap}
\begin{algorithmic}[1]
\For{$i = 1 \text{ to } B$}
\State Let $H$ be a subgraph induced on a random sample of $n$ vertices of $G_{0}$.
\State Let $P_{H}$ be the degree distribution of $H$.
\State $\hat{\theta}^{*} \gets  \underset{\theta}{\text{argmin}} \ n (P_{H} - \bm{X}P_{0}(\theta))^{\prime} \left( \text{diag}(\bm{X}P_{0}(\theta)\right) (P_{H} - \bm{X}P_{0}(\theta))$
\State $D^{*}_{i} \gets \sqrt{n} \sup | F_{H} - \tilde{F}_{0}(\hat{\theta}^{*})|$
\EndFor 
\State Let $c_{1-\alpha}$ be the $1-\alpha$ quantile of $\P_{*}(\tau \leq D^{*}_{i} )$.
\State Reject $H_{0}$ if $D(G^{\prime}) > c_{1-\alpha}$.
\end{algorithmic}
\end{algorithm}


\subsection{Sampling Graphs with a Prescribed Degree Distribution}
Let $d_{1}, \dots, d_{N}$ be the set of degrees of each vertex implied by the desired degree distribution. Begin with a set of vertices, $i = 1,\dots,N$. Endow each vertex $i$, with a set of $d_{i}$ \emph{stubs} (or \emph{half-edges}) emanating from the vertex. Now construct a random matching on the set of stubs, and connect each pair of stubs that are matched to form an edge. 

To randomly match the stubs, create a list of the vertex labels in which label $i$ appears $d_{i}$ times, then form a random permutation of the list. To construct the graph, start at the first entry in the permuted list and begin pairing off the stubs of vertices with adjacent labels. 

\begin{example}
Suppose we wish to construct a graph with degree sequence $4, 2, 2, 1, 1$.


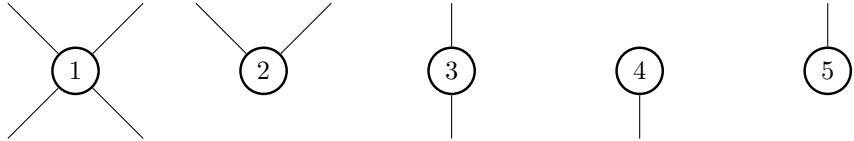
\begin{figure}[H]
\centering
\begin{tikzpicture}

\node[circle,draw,black, fill=white, line width = 0.33mm] (A) at (0,0) {$1$};
\node[circle,draw,black, fill=white, line width = 0.33mm] (B) at (\dist,0) {$2$};
\node[circle,draw,black, fill=white, line width = 0.33mm] (C) at (2*\dist,0) {$3$};
\node[circle,draw,black, fill=white, line width = 0.33mm] (D) at (3*\dist,0) {$4$};
\node[circle,draw,black, fill=white, line width = 0.33mm] (E) at (4*\dist,0) {$5$};

\coordinate (A1) at ++(\Theta:\stubside);
\coordinate (A2) at ++(180-\Theta:\stubside);
\coordinate (A3) at ++(-\Theta:\stubside);
\coordinate (A4) at ++(180+\Theta:\stubside);

\coordinate (B1) at ([shift={(\Theta:\stubside)}] B);
\coordinate (B2) at ([shift={(180-\Theta:\stubside)}] B);

\coordinate (C1) at ([shift={(90:\stubup)}] C);
\coordinate (C2) at ([shift={(3*90:\stubup)}] C);

\coordinate (D1) at ([shift={(3*90:\stubup)}] D);

\coordinate (E1) at ([shift={(90:\stubup)}] E);

\draw[color=black] (A) -- (A1);
\draw[color=black] (A) -- (A2);
\draw[color=black] (A) -- (A3);
\draw[color=black] (A) -- (A4);

\draw[color=black] (B) -- (B1);
\draw[color=black] (B) -- (B2);

\draw[color=black] (C) -- (C1);
\draw[color=black] (C) -- (C2);

\draw[color=black] (D) -- (D1);

\draw[color=black] (E) -- (E1);
\end{tikzpicture}

\caption{Vertices with 4, 2, 2, 1, and 1 stubs respectively.}
\label{fig:stubs}
\end{figure}


Construct the list: $1 1 1 1 2 2 3 3 4 5$. Produce a random permutation:

 $\mybox{51}\myboxx{13}\mybox{21}\myboxx{23}\mybox{14}$. Connect adjacent vertices.
\begin{figure}[H]
\centering
\begin{tikzpicture}

\node[circle,draw, black, fill=white, line width = 0.33mm] (A) at (0,0) {$1$};
\node[circle,draw,black, fill=white, line width = 0.33mm] (B) at (\dist,0) {$2$};
\node[circle,draw,black, fill=white, line width = 0.33mm] (C) at (2*\dist,0) {$3$};
\node[circle,draw,black, fill=white, line width = 0.33mm] (D) at (3*\dist,0) {$4$};
\node[circle,draw,black, fill=white, line width = 0.33mm] (E) at (4*\dist,0) {$5$};

\coordinate (A1) at ++(\Theta:\stubside);
\coordinate (A2) at ++(180-\Theta:\stubside);
\coordinate (A3) at ++(-\Theta:\stubside);
\coordinate (A4) at ++(180+\Theta:\stubside);

\coordinate (B1) at ([shift={(\Theta:\stubside)}] B);
\coordinate (B2) at ([shift={(180-\Theta:\stubside)}] B);

\coordinate (C1) at ([shift={(90:\stubup)}] C);
\coordinate (C2) at ([shift={(3*90:\stubup)}] C);

\coordinate (D1) at ([shift={(3*90:\stubup)}] D);

\coordinate (E1) at ([shift={(90:\stubup)}] E);

\draw[color=black] (A) -- (A1);
\draw[color=black] (A) -- (A2);
\draw[color=black] (A) -- (A3);
\draw[color=black] (A) -- (A4);

\draw[color=black] (B) -- (B1);
\draw[color=black] (B) -- (B2);

\draw[color=black] (C) -- (C1);
\draw[color=black] (C) -- (C2);

\draw[color=black] (D) -- (D1);

\draw[color=black] (E) -- (E1);

\draw[dashed, bend left] (A1) to (B2);
\draw[dashed, bend left] (A2) to (E1);
\draw[dashed, bend left] (B1) to (C1);
\draw[dashed, bend right] (A3) to (C2);
\draw[dashed, bend right] (A4) to (D1);

\end{tikzpicture}
\caption{Connect the stubs.}
\label{fig:stubs_connected}
\end{figure}
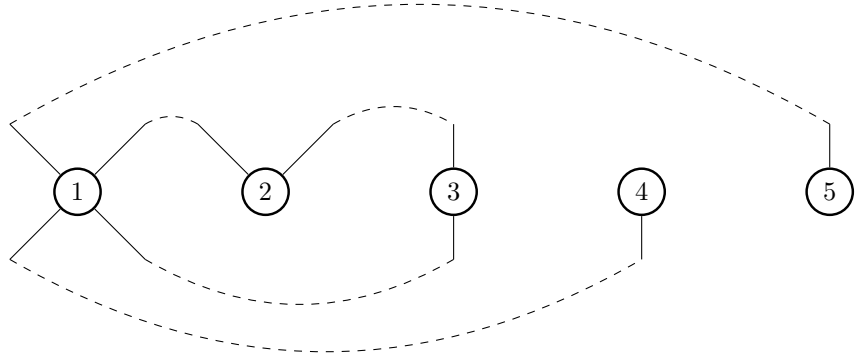

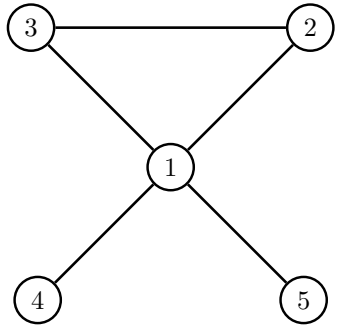
\begin{figure}[H]
\centering
\begin{tikzpicture}
\def \dist{2.5}
\node[circle,draw, black, fill=white, line width = 0.33mm] (A) at (0,0) {$1$};
\node[circle,draw,black, fill=white, line width = 0.33mm] (B) at (\Theta:1.05*\dist) {$2$};
\node[circle,draw,black, fill=white, line width = 0.33mm] (C) at (180-\Theta:1.05*\dist) {$3$};
\node[circle,draw,black, fill=white, line width = 0.33mm] (D) at (180+\Theta:\dist) {$4$};
\node[circle,draw,black, fill=white, line width = 0.33mm] (E) at (-\Theta:\dist) {$5$};

\draw[black, line width = 1] (A)--(B)
	(B)--(C)
	(A)--(C)
	(A)--(D)
	(A)--(E);

\end{tikzpicture}

\caption{Resulting Graph}
\label{fig:resultnet}
\end{figure}
\end{example} 

Given a degree sequence, $d_1, \dots, d_N$, We construct the adjacency matrix of the null graph on $N$ vertices as follows.

\begin{algorithm}[H]
\caption{Null Graph}
\label{alg:nullgraph}
\begin{algorithmic}[1]
\State Let $S$ be a vector of labels $1, \dots, N$ where label $i$ has multiplicity $d_i$
\State $A \gets zeros(N,N)$
\State $P \gets randperm(S)$
\For{$i = 1 \text{ to } N-1$ \textbf{Step} 2}
\State $A( P(i), P(i+1)) \gets 1$
\State $A( P(i+1), P(i)) \gets 1$
\EndFor 
\end{algorithmic}
\end{algorithm}

This yields an adjacency matrix $A$ from which we may sample induced subgraphs. The adjacency matrix for the subgraph induced by the set of vertices $J \subset \{j \ | \ j=1,\dots, N\}$ may be formed from extracting the set of $J$ rows and corresponding columns from $A$.

 Under this sampling method it is possible for the graph to have self-loops and multiple edges between vertices. This may either be ignored, as the probability of such an occurance is declining with the number of vertices, or the algorithm may be repeated until a simple graph is formed.

\section{Simulation Study}

We conduct a monte carlo study in order to assess the size and power properties of the Lilliefors-type test using the Kolmogorov-Smirnov test statistic. We consider testing for two common graph distributions: poisson random graph, and scale-free. The poisson random graph has a poisson degree distribution

\begin{equation}
\label{eqn:poisson}
P_{\lambda}(k) = \frac{e^{-\lambda} \lambda^{k}}{k!}
\end{equation}

and may be considered the limiting degree distribution of a graph which forms through an Erd\H{o}s-R\'{e}nyi random graph process \citep{Erdos1960} in which edges between vertices form independently with constant probability. A scale-free graph exhibits a power law distribution

\begin{equation}
\label{eqn:scalefree}
P_{\gamma}(k) = \frac{k^{-\gamma}}{\zeta(\gamma)}
\end{equation}

where $\zeta(\cdot)$ is the Riemann zeta function, and may be considered as forming through a preferential attachment process \citep{Barabasi1999, Barabasi2001}.

\begin{table}[t]
\caption{Rejection Rates. N = 10,000.}
\label{tab:rejectionrates}

\begin{tabular}{c c  ccccc}
Sampling Rate & Test Family & \multicolumn{5}{c}{Data Generating Process}\\ 
\hline\hline \\ 
&&&\multicolumn{3}{c}{Poisson} &\\\cline{4-6}
\\ 
&   	&	$\lambda = 1$ &	$\lambda = 2$ &	$\lambda = 3$ &$\lambda = 4$ &$\lambda = 5$ \\\cline{3-7}
\multirow{2}{*}{5\%} & Poisson     & 4.5 & 5.2 & 6.7 & 5.8 & 4.2\\ 
 					     & Scale-free  & 3.7 & 14.1 & 30.0 & 41.4 & 46.0\\ 
\hline 
\multirow{2}{*}{10\%} & Poisson     & 7.9 & 5.4 & 6.4 & 4.1 & 5.0\\ 
 					     & Scale-free  & 11.8 & 38.4 & 96.7 & 99.9 & 100.0\\ 
\hline 
\multirow{2}{*}{15\%} & Poisson     & 5.5 & 4.9 & 3.9 & 7.2 & 4.4\\ 
 					     & Scale-free  & 28.4 & 77.4 & 100.0 & 100.0 & 100.0\\ 
\hline 
\multirow{2}{*}{20\%} & Poisson     & 4.6 & 5.6 & 6.2 & 4.3 & 6.4\\ 
 					     & Scale-free  & 76.8 & 91.3 & 100.0 & 100.0 & 100.0\\ 
\hline 
\\ 
&&&\multicolumn{3}{c}{Scale-free} &\\\cline{4-6} 
\\ 
&   	&	$\gamma = 1.75$ &	$\gamma = 2$ &	$\gamma = 2.25$ &$\gamma = 2.5$ &$\gamma = 2.75$ \\\cline{3-7} 
\multirow{2}{*}{5\%} & Poisson     & 14.7 & 12.1 & 9.0 & 6.5 & 3.4\\ 
 					     & Scale-free  & 6.9 & 6.1 & 5.0 & 5.0 & 2.8\\ 
\hline 
\multirow{2}{*}{10\%} & Poisson     & 81.9 & 50.2 & 15.8 & 8.5 & 13.7\\ 
 					     & Scale-free  & 5.6 & 5.0 & 5.6 & 4.9 & 4.2\\ 
\hline 
\multirow{2}{*}{15\%} & Poisson     & 99.6 & 78.2 & 26.5 & 28.7 & 55.4\\ 
 					     & Scale-free  & 5.9 & 4.7 & 5.2 & 6.5 & 4.8\\ 
\hline 
\multirow{2}{*}{20\%} & Poisson     & 100.0 & 99.8 & 85.3 & 72.1 & 90.9\\ 
 					     & Scale-free  & 4.9 & 4.4 & 4.8 & 4.5 & 3.9\\ 
\hline 
\hline\hline 
\end{tabular}
\end{table}

\begin{table}[t]
\caption{Rejection Rates. N = 20,000.}
\label{tab:rejectionrates2}
\begin{tabular}{c c  ccccc}
Sampling Rate & Test Family & \multicolumn{5}{c}{Data Generating Process}\\ 
\hline\hline \\ 
&&&\multicolumn{3}{c}{Poisson} &\\\cline{4-6}
\\ 
&   	&	$\lambda = 1$ &	$\lambda = 2$ &	$\lambda = 3$ &$\lambda = 4$ &$\lambda = 5$ \\\cline{3-7}
\multirow{2}{*}{5\%} & Poisson     & 3.3 & 5.9 & 4.5 & 5.0 & 4.6\\ 
 					     & Scale-free  & 5.9 & 20.3 & 51.3 & 69.9 & 83.0\\ 
\hline 
\multirow{2}{*}{10\%} & Poisson     & 6.8 & 5.6 & 6.0 & 4.9 & 5.8\\ 
 					     & Scale-free  & 14.2 & 77.1 & 100.0 & 100.0 & 100.0\\ 
\hline 
\multirow{2}{*}{15\%} & Poisson     & 5.4 & 5.1 & 4.7 & 5.1 & 5.6\\ 
 					     & Scale-free  & 42.8 & 95.8 & 100.0 & 100.0 & 100.0\\ 
\hline 
\multirow{2}{*}{20\%} & Poisson     & 6.4 & 5.1 & 4.7 & 3.9 & 5.4\\ 
 					     & Scale-free  & 99.7 & 99.9 & 100.0 & 100.0 & 100.0\\ 
\hline 
\\ 
&&&\multicolumn{3}{c}{Scale-free} &\\\cline{4-6} 
\\ 
&   	&	$\gamma = 1.75$ &	$\gamma = 2$ &	$\gamma = 2.25$ &$\gamma = 2.5$ &$\gamma = 2.75$ \\\cline{3-7} 
\multirow{2}{*}{5\%} & Poisson     & 23.5 & 6.8 & 5.5 & 7.9 & 6.0\\ 
 					     & Scale-free  & 4.3 & 5.6 & 4.8 & 6.1 & 4.0\\ 
\hline 
\multirow{2}{*}{10\%} & Poisson     & 97.3 & 80.7 & 33.4 & 11.6 & 24.5\\ 
 					     & Scale-free  & 5.5 & 4.5 & 4.6 & 4.1 & 5.1\\ 
\hline 
\multirow{2}{*}{15\%} & Poisson     & 100.0 & 99.8 & 65.3 & 32.0 & 70.4\\ 
 					     & Scale-free  & 5.4 & 5.7 & 4.8 & 3.8 & 5.5\\ 
\hline 
\multirow{2}{*}{20\%} & Poisson     & 100.0 & 100.0 & 100.0 & 98.3 & 99.9\\ 
 					     & Scale-free  & 5.2 & 5.3 & 4.4 & 5.2 & 4.4\\ 
\hline 
\hline\hline 
\end{tabular}
\end{table}

The data generating process (DGP) in each simulation consists of randomly selecting $n \in \{500, 1000, 1500, 2000\}$ vertices from an underlying population graph on $N=10,000$ vertices and sampling the subgraph induced by those vertices. We consider underlying population graphs in the poisson family with parameters $\lambda \in \{1,2,3,4,5\}$, and the scale-free family with $\gamma \in \{1.75, 2, 2.25, 2.5, 2.75\}$. Adjacency matrices for the underlying population graphs for each parameter are generated by \cref{alg:nullgraph}.

For each DGP we test the hypotheses that the data were drawn from poisson and scale-free distributions respectively such that in each simulation we test a null hypothesis which is true (size) and one which is false (power). For the test of each hypothesis we employ the bootstrap in \cref{alg:bootstrap} to extract critical values with 500 bootstrap pseudo-samples. \Cref{tab:rejectionrates} displays the rejection frequencies over 1000 iterations for each DGP.

We see that for both DGPs the bootstrap is able to control the size well, keeping the rejection frequencies close to the nominal rate of 5\%. For the poisson DGP, power is increasing in the rate parameter. This is expected, as the larger the rate parameter, the more mass is pushed away from low degree vertices compared with scale-free. For rate parameters close to 1, the poisson and scale-free distributions are practically indistinguishable, with almost all of their mass in low degree vertices. For the scale-free DGP, power is large for low scale-free parameters, which produces fatter tails. For both DGPs, power is increasing in the sampling rate.

We repeated the experiment with a graph on $N = 20,000$ vertices for sample sizes $n\in \{1000, 2000, 3000, 4000\}$, so that the effect of sample size on power may be determined while holding the sampling rate fixed. \Cref{tab:rejectionrates2} displays these results which show that power is in fact increasing in sample size alone.


\section{Application}

\subsection{Protein interaction networks}

A binary protein interaction network is a graph in which vertices represent the set of proteins within a cell or organism's proteome with an edge present between two vertices if the corresponding proteins are able to bind to one another. The degree distribution of a PIN is a particularly interesting feature of the graph as the degree of a protein in a PIN has been linked to its essentiality \citep{Jeong2001}. That is, the consequences of deleting genes which express low degree proteins on the viability of a cell are relatively small compared to the deletion of genes which express high degree proteins. Supposing a PIN followed a scale-free distribution, then it would inherit the associated robust-yet-fragile properties \citep{Albert2000, Doyle2005}, wherein the network is robust to failures of random vertices but vulnerable to the targeted attack of high degree vertices. \citet{Winzeler1999} and \citet{Ross-Macdonald1999} have demonstrated such resilience to random gene deletion in the budding yeast \emph{Saccharomyces cerevisiae}. In this case, proteins of high degree would present a target for therapeutic mediation, such as the p53 tumor-suppressor protein \citep{Vogelstein2000}. However, if the true degree distribution of a PIN were say, Erd\H{o}s-R\'{e}nyi, then such a strategy would not be viable.

One of the more popular methods for sampling PINs is the yeast two-hybrid assay (Y2H), due to \citet{Fields1989}. Y2H checks for the binary interaction between two proteins, named the \emph{bait} and \emph{prey}, and is performed \emph{in vivo} within an S. \emph{cerevisiae} cell. The bait is expressed with a GAL4 DNA binding domain, and the prey with a GAL4 activation domain, which allow the proteins to be attached to binding sites on the yeast's genome. If the bait and prey interact, an entire GAL4 transcription factor is reconstituted and causes the expression of a reporter gene downstream of the GAL4 binding site which allows the interaction to be detected. High-throughput Y2H assays allow for thousands of protein pairs to be tested. Since edges are being detected only between proteins being sampled, this design results in an induced subgraph sample.

\subsection{The degree distribution and measurement error}


\begin{figure}[t]
\begin{subfigure}[b]{0.45\textwidth}
\includegraphics[width=\textwidth]{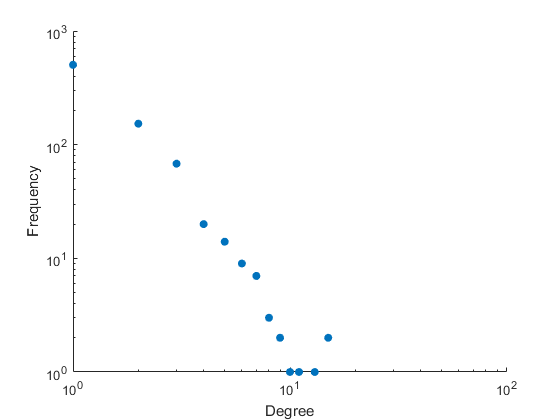}
\caption{Ito-core}
\label{subfig:itocore}
\end{subfigure}
\hfill
\begin{subfigure}[b]{0.45\textwidth}
\includegraphics[width=\textwidth]{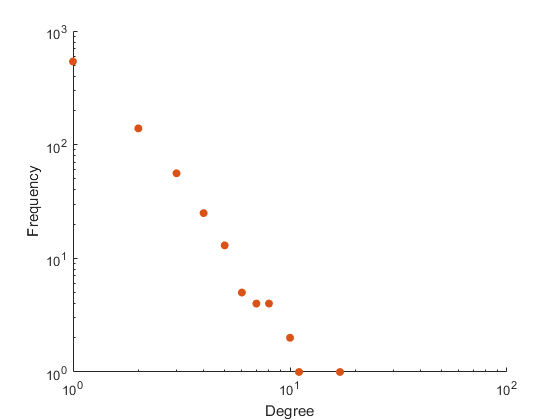}
\caption{Uetz-screen}
\label{subfig:uetzscreen}
\end{subfigure}
\caption{Empirical degree distributions for S. \emph{cerevisiae} PIN on log-log scale.}
\label{fig:empiricaldegdist}
\end{figure}


\Cref{fig:empiricaldegdist} displays the empirical degree distributions of the Ito-core \citep{Ito2001} and Uetz-screen \citep{Uetz2000} Y2H assays of the S. \emph{cerevisiae} PIN. When plotted on a log-log scale, these exhibit the characteristic straight line of a scale-free distribution. However, caution must be taken as these data sets constitute samples of size 813 and 437 proteins respectively from an estimated proteome size of 6000 \citep{Goffeau1996}. \citet{Stumpf2005a} and \citet{Stumpf2005b} show that subgraphs of scale-free networks are \emph{not} scale-free, so the apparent power law distributions in \cref{fig:empiricaldegdist} do not offer compelling evidence that the yeast PIN is scale-free. Further compounding the problem of making inferences from Y2H samples is the prevalence of false positives and false negatives, with false negative rates being as high as 90\% \citep{Hart2006, Ito2002}. \citet{Han2005} show via simulation study that when measurement error is introduced, graphs with Erd\H{o}s-R\'{e}nyi, scale-free, and exponential distributions can lead to sampled subgraphs whose degree distributions are practically indistinguishable from one another. In what follows we consider conducting inference on the yeast PIN explicitly accounting for the sampling and false negative rates.

\subsection{Data}

We consider five data sets on the S. \emph{cerevisiae} PIN taken from \citet{Yu2008}, available for download from the Center for Cancer Systems Biology yeast interactome database.\footnote{Available at \url{http://interactome.dfci.harvard.edu/S_cerevisiae/index.php?page=home}} This includes the two sets displayed in \cref{fig:empiricaldegdist}: Ito-core \citep{Ito2001}, consisting of 843 interactions between 813 proteins; and Uetz-screen \citep{Uetz2000}, with 682 interactions between 806 proteins. \citet{Yu2008} test the quality of these data sets by comparing them to two independent assays and find that both data sets exhibit high specificity, with the interactions detected likely to be true positives, but low sensitivity. \citet{Yu2008} also construct three additional data sets: a binary gold standard (Binary-GS) set of 1,318 high-confidence interactions on 1,090 proteins; the CCSB-YI1 set consisting of 1,809 interactions among 1,278 proteins with specificity at least 94\%; and Y2H-Union, which is the union of Ito-core, Uetz-screen, and CCSB-YI1. Y2H-Union is estimated to be 20\% sensitive.

\subsection{Method and Results}


\begin{table}[t] 
\begin{adjustbox}{width=1\textwidth, center}
\centering 
\begin{tabular}{c  c  cc  c cc  c cc} 
Data Set & Test Family & \multicolumn{8}{c}{False Negative Rate}\\\hline\hline 
\\ 
&& \multicolumn{2}{c}{70\%} && \multicolumn{2}{c}{80\%} && \multicolumn{2}{c}{90\%}\\\cline{3-4}\cline{6-7}\cline{9-10} 
\\ 
&& $\hat{\theta}$ & P-value	&& $\hat{\theta}$ & P-value	&& $\hat{\theta}$ & P-value\\ 
\multirow{3}{*}{Ito-core}	& Poisson & 19.39 & 0.00 && 29.31 & 0.00 && 59.03 & 0.00 \\ 
                              & Scale-free & 1.74 & 0.18 && 1.61 & 0.48 && 1.29 & 0.18 \\ 
                          	& Exponential & 0.07 & 0.05 && 0.05 & 0.07 && 0.02 & 0.04 \\ 
\hline 
\Tstrut 
\multirow{3}{*}{Uetz-screen}	& Poisson & 16.30 & 0.00 && 24.67 & 0.00 && 49.75 & 0.00 \\ 
                              & Scale-free & 1.80 & 0.49 && 1.68 & 0.87 && 1.39 & 0.49 \\ 
                          	& Exponential & 0.08 & 0.11 && 0.05 & 0.11 && 0.03 & 0.06 \\ 
\hline 
\Tstrut 
\multirow{3}{*}{CCSB-YI1}	& Poisson & 13.51 & 0.00 && 20.36 & 0.00 && 40.90 & 0.00 \\ 
                              & Scale-free & 1.62 & 0.31 && 1.50 & 0.65 && 1.31 & 0.00 \\ 
                          	& Exponential & 0.07 & 0.00 && 0.05 & 0.00 && 0.03 & 0.00 \\ 
\hline 
\Tstrut 
\multirow{3}{*}{Binary-GS}	& Poisson & 28.03 & 0.00 && 42.22 & 0.00 && 84.77 & 0.00 \\ 
                              & Scale-free & 1.59 & 0.00 && 1.39 & 0.00 && 0.84 & 0.02 \\ 
                          	& Exponential & 0.05 & 0.02 && 0.03 & 0.03 && 0.01 & 0.01 \\ 
\hline 
\Tstrut 
\multirow{3}{*}{Y2H-Union}	& Poisson & 11.07 & 0.00 && 16.68 & 0.00 && 33.47 & 0.00 \\ 
                              & Scale-free & 1.72 & 0.00 && 1.58 & 0.00 && 1.35 & 0.17 \\ 
                          	& Exponential & 0.09 & 0.00 && 0.06 & 0.00 && 0.03 & 0.00 \\ 
\hline\hline 
\end{tabular} 
\caption{Parameter estimates and p-values corresponding to the hypothesis test for each data set belonging to a poisson, scale-free, and exponential family of distributions respectively. Each test is conducted assuming false negative rates of 70\%, 80\%, and 90\% respectively, with 500 bootstrap samples.} 
\label{tab:results} 
\end{adjustbox}
\end{table}


First we must address the distortion to the degree distribution imposed by sampling an induced subgraph with the possibility of false negatives. For simplicity we will assume that vertices are chosen through Bernoulli sampling, in which each vertex is included in the sample independently with probability $p$. Assume further that for any edge present in the population graph, that edge may only be observed in an induced subgraph independently with probability $r$. Thus if we consider a vertex in the sample with $k$ edges in the population, the probability it has $i$ edges in the induced subgraph is $Binom(rp, k)$. For small $rp$ this may be well approximated as $Poiss(rpk)$. Therefore we construct the design matrix in equation \ref{eqn:distortion} such that its columns are poisson.

\begin{equation}
\mathbf{X}_{i,k+1} = \frac{e^{-rpk} (rpk)^{i}}{i!} \quad k=0,\dots, n-1
\end{equation}

Following \citep{Goffeau1996} we set $N = 6,000$ to be the size of the yeast proteome. We set $p = \frac{n}{N}$, where $n$ is the sample size of the corresponding data set. We consider false negative rates of 70\%, 80\%, and 90\%, corresponding to $r \in \{.3, .2, .1\}$. For each data set and false negative rate, we test three hypotheses: that the data were drawn from a graph whose degree distribution is poisson, $P_{G}(k; \theta) \propto \theta^{k}$; a graph whose degree distribution is scale-free, $P_{G}(k; \theta) \propto k^{-\theta}$; and a graph  whose degree distribution is exponential, $P_{G}(k; \theta) \propto e^{- \theta k}$. Each test is performed at the 5\% significance level with $B = 500$ bootstrap samples.

Adjacency matrices for the sample graphs were constructed using data from the CCSB interactome database mentioned above. Self-loops have been removed such that each graph is simple and undirected. The data do not contain information on the number of isolate vertices measured. As such, our estimation procedure and bootstrap have been modified to mimic this by removing isolates from consideration.

\Cref{tab:results} displays the results. The immediate result is that for each data set and each false negative rate, the hypothesis that the yeast PIN follows a poisson distribution is unanimously rejected. For Ito-core and Uetz-screen, only the scale-free hypothesis is retained. Scale-free is the only hypothesis retained for the CCSB-YI1 data set, except for at a false negative rate of 90\% for which all hypotheses are rejected. Y2H-Union rejects all hypotheses except for scale-free at the 90\% false negative rate. This is seems odd considering Y2H-Union is the combination of Ito-core, Uetz-screen, and CCSB-YI1; all of which are unable to reject at the 70\% and 80\% false negative levels. Binary-GS is especially odd, only retaining the exponential family for the 70\% and 80\% rates while switching to only retain scale-free at the 90\% rate. The fact that the parameter estimates for the Binary-GS data set is far out of line with those of the other sets suggests this may be an artifact of how the data were aggregated. 

These results need to be taken with a grain of salt. It is not necessarily true that the sampling methodology may be well approximated by random sampling of proteins within the S. \emph{cerevisiae} proteome. Further the rates of false positives and negatives are not known with precision. What we do glean from these results is that the Erd\H{o}s-R\'{e}nyi random graph model, resulting in poisson degree distributions, is not a good candidate for the degree distribution of the S. \emph{cerevisiae} protein interaction network. \\


\section{Discussion}

We have proposed a procedure for conducting \newline goodness-of-fit tests for sampled subgraphs. The procedure relies on a parametric bootstrap being adapted to mimic the subgraph sampling process. A monte carlo study reveals that this test posesses good size properties, keeping the type I error rate close to the specified nominal level, while having good power in a large region of the parameter space. In this paper we have focused on a sampling design where a subset of vertices are selected, either by Bernoulli or simple random sampling, and the subgraph induced by those vertices constitutes our sample; though this is not the only sampling design servicable by our methodology. It applies to any sampling design where the distortion to the degree distribution may be modelled as a linear transformation, as in equation \ref{eqn:distortion}, where the matrix $\mathbf{X}$ does not depend on features of the graph being sampled. Incident edge, traceroute, snowball, and ego sampling are all included in this subset. 

A possible impediment to the utility of this approach is that for the bootstrap to function, a graph on $N$ vertices must be constructed, where $N$ is the number of vertices in the population. For large online social networks, which can contain millions of vertices and edges, storage of the corresponding adjacency matrix, and the time required to sample subgraphs from it, may be too burdensome. An efficienct implementation of the bootstrap for such large graphs presents an area for future research. \\

\clearpage
\bibliographystyle{apalike}
\bibliography{AGOFTSS}


\end{document}